\documentclass[fleqn,twoside]{article}
\usepackage{espcrc2}

\usepackage{graphicx}
\usepackage[figuresright]{rotating}

\newcommand{\AmS}{{\protect\the\textfont2
  A\kern-.1667em\lower.5ex\hbox{M}\kern-.125emS}}

\title{Lightest Nuclei in UHECR versus Tau  Neutrino Astronomy }

\author{D.Fargion\address[MCSD]{Physics Department and INFN, Rome University 1, Sapienza, Ple.A.Moro 2}, D. D'Armiento, P. Paggi, S. Patri'
        }

\begin{document}

\begin{abstract}
UHECR may be either nucleons \cite{Auger-Nov07} or nuclei; in the latter case the Lightest Nuclei, as $He^4$ or $He^3$, Li, Be , explains at best the absence of Virgo signals and the crowding of events around Cen-A bent by galactic magnetic fields \cite{Fargion2008}. This model fit the observed nuclear mass composition discovered in AUGER. However UHECR nucleons above GZK produce EeV neutrinos while Heavy Nuclei, as Fe UHECR do not produce much. UHECR He nuclei at few $10^{19} eV$ suffer nuclear fragmentation (producing low energetic neutrino at tens PeVs) but it  suffer anyway  photo-pion GZK suppression (leading to EeV neutrinos) once above  one-few $10^{20} eV$. Both these cosmogenic UHE secondary neutrinos signals may influence usual predicted GZK  \cite{Greisen:1966jv},\cite{za66}   Tau Neutrino Astronomy \cite{Fargion1999} in significant and detectable way; the role of resonant antineutrino electron-electron leading to Tau air-shower may also rise.
\vspace{1pc}
\end{abstract}

\maketitle

\section{ UHECR  Lightest Nuclei versus UHE $\nu$}
\begin{figure}[htb] \vspace{12pt}
\includegraphics[width=60mm]{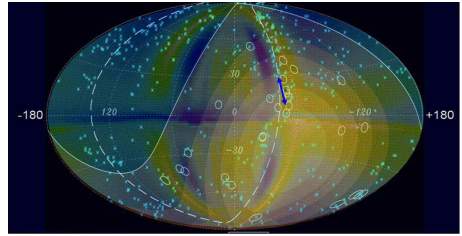}
\caption{The first order bending for lightest UHECR, as He nuclei, is shown by the vertical arrow. The underline
galactic magnetic fields almost vertical to Galactic Plane in Cen-A region,  explain the longitudinal
clustering of the events, overlapping by chance on Super Galactic Arm , shown by the dashed curve. It also explain the UHECR composition and the absence of Virgo. The heavier, more charged and more energetic ($He^4$,Li,Be) will be bent more. The lighter and less energetic ${He^3}$ less. } \label{UHECR-MAP}
\end{figure}

Astrophysical UHE neutrinos is being searched, since four decades, at TeV-PeV energy via neutrino muons in underground detectors. Their signal is greatly polluted from above (and below) by CR secondary muons (and atmospheric muon neutrinos into muons) and partially suppressed at PeVs, by Earth size opacity. Horizontal Muons at hundreds TeV, originated by astrophysical UHE neutrinos are also  polluted by leading prompt atmospheric signals. However, since four decades, the rise of highest energy neutrino Astronomy is generally expected also by the  cosmogenic EeV  neutrinos, secondaries of  GZK \cite{Greisen:1966jv},\cite{za66} cut-off on Ultra High Cosmic Rays (UHECR). These ones are difficult to observe in $km^3$ detector. But since a decade \cite{Fargion1999} the UHE Tau Neutrino Astronomy have been foreseen and suggested in present and future detectors.  Indeed Tau Air-shower Astronomy is an amplified Neutrino Astronomy observable. Indeed an air-shower is not a single track, but a wide (even  many $km^2$) area raining of billions secondaries at once.  One doesn't need to collect and count all of them to realize such an event beyond a mountain or Earth screen. One need just a spread sample at once. As for UHECR.  Moreover hadronic air-shower below  zenith angle $75^o$ are originated far and far and they are filtered : their peak electromagnetic component is exhausted at  horizons leaving just late diluted (and sharp) inclined muon bundles. On the contrary neutrino, either interacting in air or, better, escaping the Earth by $\nu_{\tau}\rightarrow \tau$ may lead to electromagnetic-rich horizontal (upward) air-showers. Tau decay in flight are better than neutrino interacting in air  because Earth rock density is three thousand times larger than air one. Even if the tau enjoy of a bounded escaping solid angle (a zenith angle width of $2-5^o$ degree) while all down-ward neutrinos interacting on air may reach from wider cone (a zenith angle width  $15^o$ degree) and are in three flavors. Such skimming events in AUGER experiment may rise via upward tau air-shower \cite{Fargion1999},\cite{Bertou2002},\cite{Bigas07}. In last years the upward-horizontal EeV $\tau$,$\bar{\tau}$ appearance, via UHECR $p + \gamma_{CMB}\rightarrow \pi \rightarrow \nu $  has been predicted by many authors; the most extreme ones were at rate of $0.1-0.03$ a year\cite{Bigas07}, or  $0.3$ a year in AUGER \cite{Fargion2007} and finally up to $0.2$ a year \cite{Auger08}. This rate, has only recently being adjusted and confirmed by last AUGER group estimates (NOW 2008, CRIS 2008):$0.3$ a year, in full agreement with our previous ones \cite{Fargion2007}. Indeed following the   AUGER evidences (and Hires ones \cite{Hires06})  of  an UHECR GZK\cite{Greisen:1966jv},\cite{za66} cut-off and the latest AUGER (possible) Super-Galactic anisotropy due to an eventual proton UHECR guarantee a secondary flux of UHE-GZK neutrino at EeVs energy within AUGER detection via $\tau$,$\bar{\tau}$ showering \cite{Fargion2007}. In fact this occurs because the muon neutrino flavor mixing must feed also a  tau neutrino component. Such UHE astrophysical tau neutrino (noise-free  from any atmospheric background) may interact in and it may rises out the Earth as UHE $\tau$. The UHE $\tau$,$\bar{\tau}$ decay in flight in atmosphere must lead to loud Tau Air-showers. Such a detectable flashes may rise in short times within Auger SD (by large electromagnetic curvature signals) or in FD arrays by horizontal fluorescence signals, namely once in a few years ($2-4$) from now \cite{Fargion2007}. Nevertheless a recent alternative UHECR understanding \cite{Fargion2008}, based on observed AUGER UHECR  (nuclei) mass composition and with Cen-A rich clustering map, is in disagreement with UHECR proton understanding \cite{Auger-Nov07}. This model is leading to different UHE neutrino  predictions. It suggests that UHECR are made by Lightest Nuclei ($He^4$, $He^3$, maybe a few also $Be$,$Li$) mostly originated from Cen-A: their trajectories are bent and spread by galactic magnetic fields and they are  incidentally clustered  ( by galactic fields) around Cen-A, nearest ($4$ Mpc) and possibly unique source able to survive the short Lightest Nuclei GZK cut-off. These events spread mostly along the same  Super-Galactic Arm , just apparently from far $80$ Mpc  Centaurs Cluster. The mean random angle  bending $He^4$ (see Fig.\ref{UHECR-MAP}) by spiral galactic magnetic  fields along the plane is $\delta_{rm} \geq$:
\begin{equation}
{11.3^\circ}\cdot \frac{Z}{Z_{He^4}} \cdot (\frac{6\cdot10^{19}eV}{E_{CR}})(\frac{B}{3\cdot \mu G})\sqrt{\frac{L}{20 kpc}}
\sqrt{\frac{l_c}{kpc}}
\end{equation}

This {\emph{Lightest Nuclei for Highest Cosmic Rays}} model  implies and foresees among the other, additional clustering of UHECR events around the nearest AGN Cen-A (the lightest UHECR the more correlated to the source, the heavier and with larger charges, the most bent and spread ones). The model        explains the absence or a poor  signal  from Virgo (too far for the fragile nuclei to fly by). Such {\emph{Lightest Nuclei for Highest Cosmic Rays}}  are forced in a very confined cosmic volume (ten Mpc) due to a fragile light nuclear (few MeVs) binding energy. Usually heavy nuclei fragmentation pour energy only  in UHE neutrino (at $0.1-0.01$ EeV energy) spectra, less energetic than common UHE EeV $p+\gamma_{CMB}\rightarrow\pi$ neutrino flux.

\section{UHECR by Lightest Nuclei versus $\nu_{\tau}$ }
\begin{figure}[htb] \vspace{9pt}
\includegraphics[width=60mm]{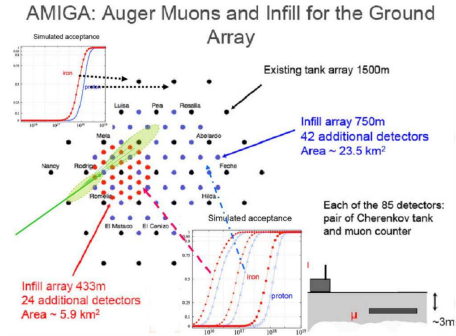}
\caption{ The AUGER Array and the inner area AMIGA and HEAT where lower energy air-showers might be revealed.
The two different spacing (5.9 $km^2$ and the 23.5 $km^2$) offer a lower energy threshold for neutrino inclined air-showers respectively at $10^{16}$ and $10^{17}$ eV . In the figure a schematic area due to air-showering lobes of an escaping horizontal air-shower } \label{Array}
\end{figure}

AUGER SD or FD are not able to reveal tens-hundreds PeV energy  easily. However a very recent and a less spaced AUGER sub-system, a more dense array AMIGA and the additional telescopes HEATS might lower the threshold accordingly. The $0.1 EeV$ mostly hadronic, inclined-\emph{upward}, tau air-showers $\theta \leq 80^o$ occur at much lower altitudes than hadronic inclined \emph{down}-going air-showers,
at much nearer distances from telescopes than hadronic EeV air-showers, reducing the area and the rate. They may offer a tens-hundred PeV neutrino windows, secondaries of UHE He nuclear fragmentation. Some estimates are  offered assuming, for sake of simplicity a Fermi-like
UHE GZK spectra, comparable to the well known Waxmann-Bachall flux derived for cosmic GRBs.  Nevertheless if rarest UHECR, possibly from AGN,  above $1-2 \cdot 10^{20}  eV$, ( a few in AUGER events), are also He nuclei, they may still suffer GZK photo-pions opacity decaying into EeV neutrinos too. Such He GZK interaction  may be comparable with corresponding proton GZK ones at $6 \cdot 10^{19} eV$: therefore EeV tau Neutrino Astronomy, tails of the most energetic Lightest Nuclei GZK secondaries, may still rise at AUGER.  EeV inclined tau decays, born nearly at sea level (versus inclined hadronic $\theta \geq 80^o$ ones developing at altitudes well above ten km) occur at air-density  at least three times higher (than hadron horizontal ones); therefore  Tau Air-showers (mostly behaving as hadron ones), their electromagnetic and fluorescence tail  sizes (and times) are three times shorter ($l_{sh}\approx8-10$ km) and their characteristic azimuth speed  $\dot{\varphi}\approx 1.5 \cdot 10^4 rad s^{-1}$ is slower than common inclined UHECR hadron ones at higher altitudes ($l_{sh}\geq 25-30$ km), ($25-35$ km), whose detection threshold (minimal impact distance) is very near to the observer (a dozen of km from the edges and few kms from air-shower baricenter). The consequent azimuth angular velocity is for instance, assuming an EeV event, at maximal distance of $12$ km, at minimal distance $9$ km, total duration 100 $\mu$ s,  $\dot{\varphi}\geq 2.1 \cdot 10^4 rad s^{-1}$. Such a brief $26-30\mu$s and distinct  up-ward tau lightening versus slow $90-120 \mu$s diluted down-ward horizontal showering  may be  well disentangled within AUGER (and AMIGA SD) threshold, also by obvious angular resolution and directionality (up or downward); their strong curvature and their inclinations would rise within HEAT FD and AUGER FD, we estimate, once every a couple of years \ref{Array}. Possibly (factor two) originated from the  Ande side (West to East )\cite{Fargion1999},\cite{aramo05},\cite{Miele06}.

\begin{figure}[htb] \vspace{9pt}
\includegraphics[width=80mm]{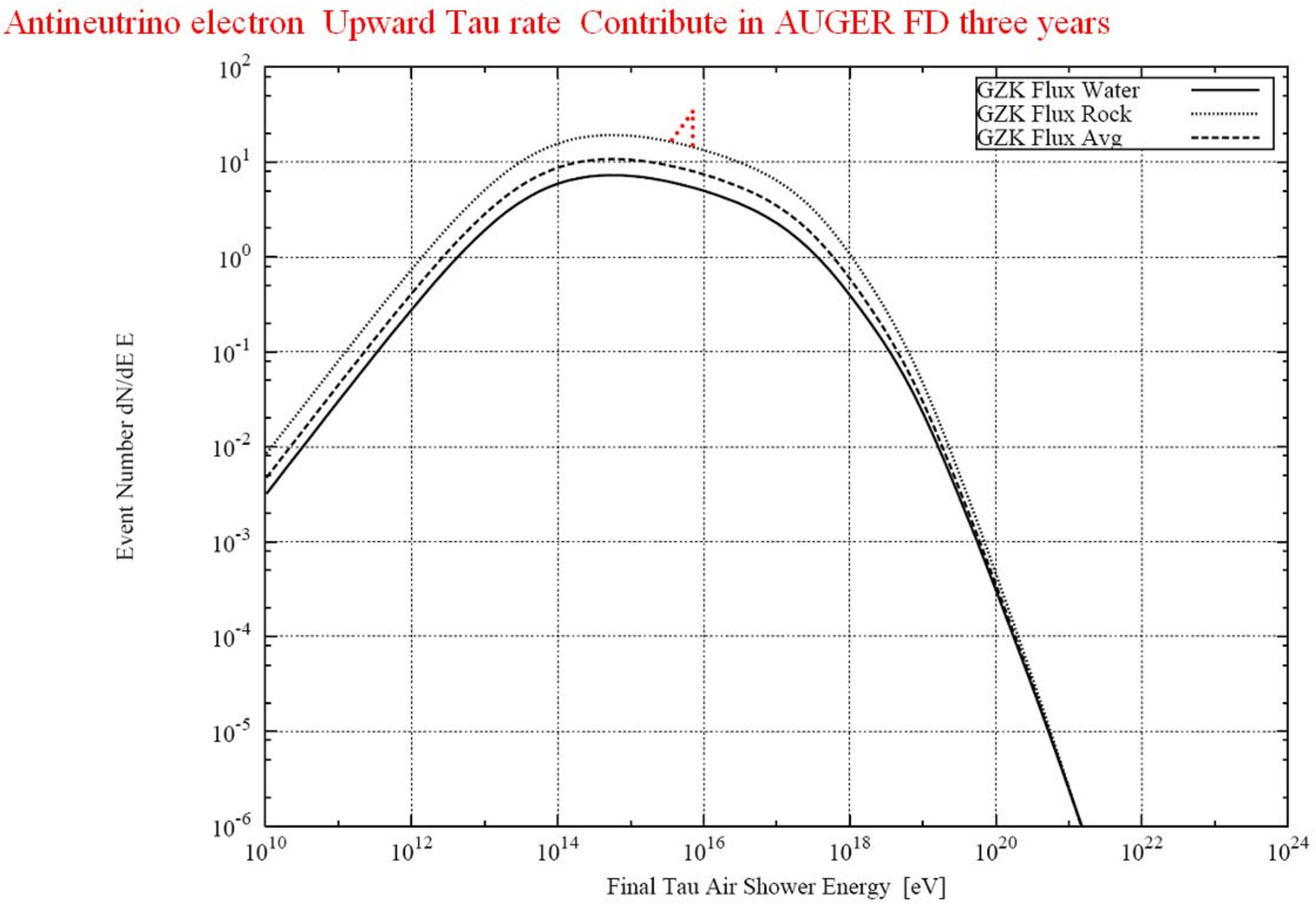}
\caption{ As above the integral rate of up going Tau including the resonant neutrino contribute in three years. The small triangular
 bump due to Glashow resonant antineutrinos scattering on electrons is mainly due to the energy spread of the tau energy at its  birth. The rate are estimated for a total AUGER area at  10$\%$ efficiency for FD. The FD threshold is growing linearly with energy and it is respectively $300 Km^2$, $30 Km^2$, $3 Km^2$ at the energy $EeV$, $0.1 EeV$, $0.01$ EeV . Because of it each rate at lower energies than EeV must be suppressed in FD efficiency by corresponding factor (0.1, 0.01) making the expected event rate at ten PeV ($0.26$) and hundred PeV (0.5), just  below unity in 3 years. The additional mini-array AMIGA at $27.5 Km^2$, $6 Km^2$, whose array spacing is respectively ,$750$, $433$ m. is a SD active day and night and it might double the signal, offering a detection in a few   years from now  (0.35-0.4 event/year).} \label{Integral}
\end{figure}

This new neutrino Astronomy, the PeVs-EeVs ratio, may disentangle in future records  the real UHECR  nucleon or  lightest nuclei UHECR nature. Indeed the additional rise  of the resonant Glashow  contribute, $\bar{\nu}_e+ e \rightarrow W^-\rightarrow \bar{\nu}_{\tau} + \tau $, in the upgoing Tau, while just marginally doubling the signal at 0.01 EeV , see Fig.\ref{NuLenght},\ref{Differential},\ref{Integral} it may leave a clear imprint on horizontal  differential angular  spectra. Leading to a flavor UHE neutrino spectroscopy.
 \subsection{The resonant $\bar{\nu}_e+ e \rightarrow W^-\rightarrow \bar{\nu}_{\tau} + \tau $}
\begin{figure}[htb] \vspace{9pt}
\includegraphics[width=60mm]{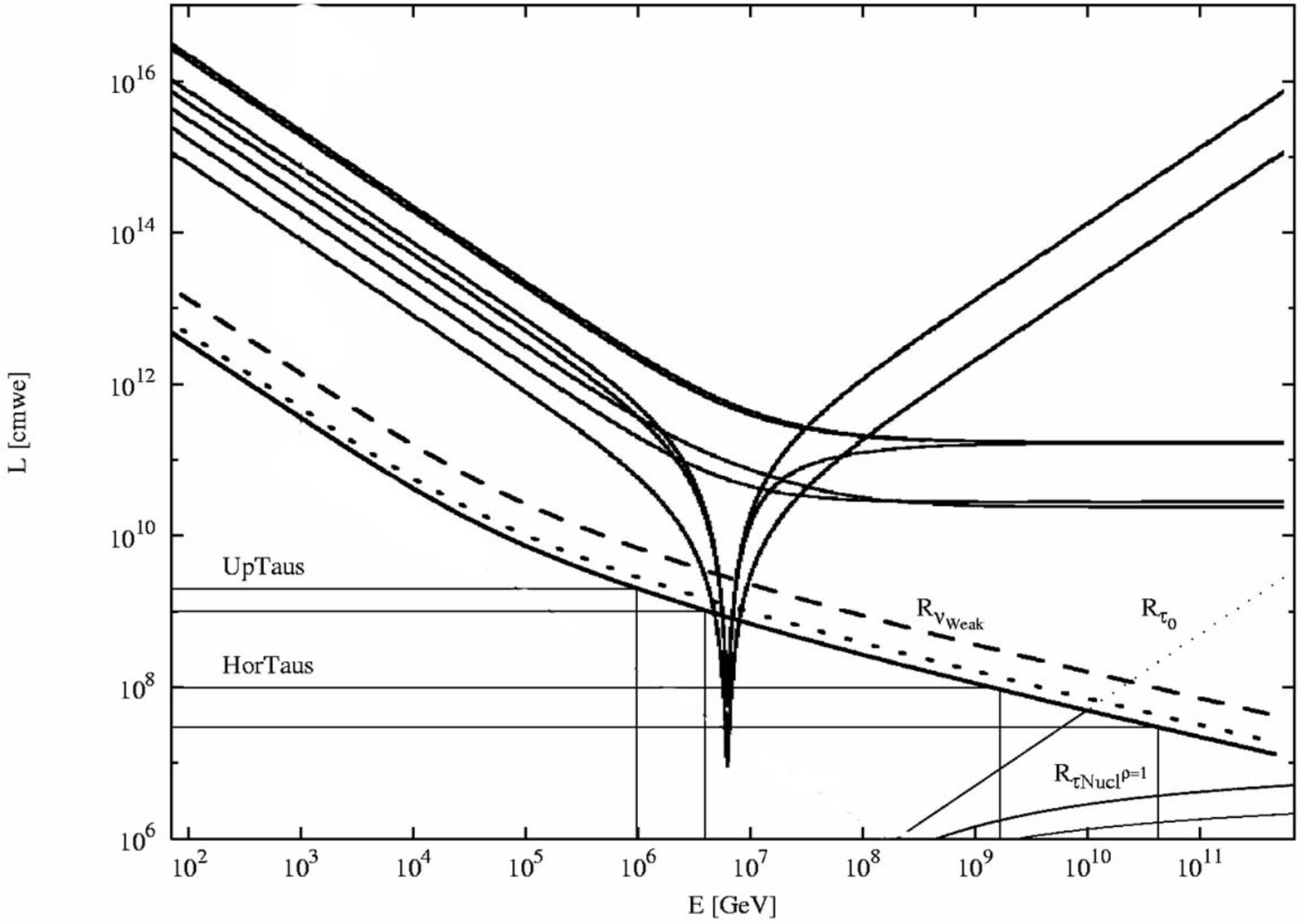}
\caption{The neutrino interaction length either with nuclear matter as well as for anti-neutrino $\bar{\nu}_e$  resonant with electron. The two horizontal roads UpTaus and HorTaus define the corresponding energy for Upward and horizontal showers from Earth \cite{Fargion1999},\cite{Fargion2007}. The Glashow resonant energy (inverse peak) corresponds to the EeV neutrino-nucleon interaction length: Note  the average peak interaction  with EeV energy one.} \label{NuLenght}

\includegraphics[width=72mm]{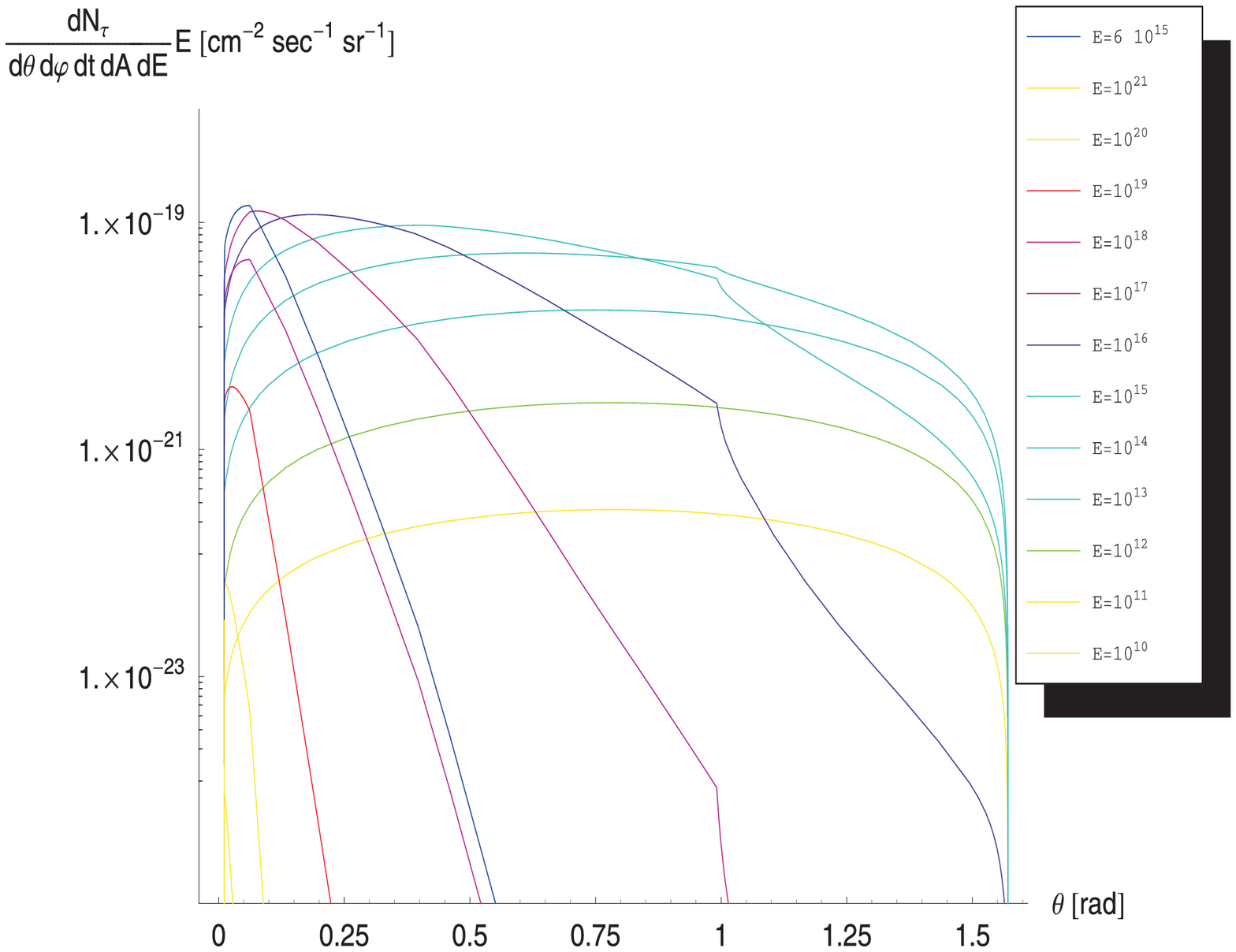}
\caption{The differential angular distribution of upgoing Tau assuming an usual Waxmann Bachall neutrino energy fluency of
$10 eV \cdot cm^{-2} \cdot s^{-1} \cdot sr^{-1}$ , for each neutrino flavor, derived from the observed UHECR GZK\cite{Greisen:1966jv},\cite{za66} at $6 \cdot 10^{19}$ eV. The inner Earth density profile has been taken into account. Note the very peculiar bump due to the resonant antineutrino at $6.3$ PeV; its preferential inclination is due to the severe  Earth opacity. The overall integral signal almost double the Tau skimming rate in the same energy band} \label{Differential}
\end{figure}

   The role of UHE neutrino interaction with matter is well understood: they also shape the neutrino   survival across the Earth. Indeed the highest energy $\nu$ are opaque to Earth, but not to   smallest cord. Therefore the harder events are the more opaque and survive at shortest cords, the tangent ones. The
    resonant $\bar{\nu_e} + e \rightarrow W^-\rightarrow \bar{\nu_{\tau} + \tau}$ are very peculiar    signals. Their opacity on Earth is extreme. They are as opaque as EeV energetic neutrinos.See \ref{NuLenght}.
    But their lower energy corresponds to higher flux (for constant energy fluency as Waxmann Bachall one).    But their propagation in Earth is much smaller too. The compensated flux in 5-6 times higher than EeV one.
    Unfortunately this rate is (in AUGER) suppressed by detection threshold (nearly one hundred times smaller    than EeV showers). Nevertheless the mini array AMIGA (of 6 $km^2$) and the HEAT telescopes may contribute
    to make detectable in a few year the same resonant  $\bar{\nu_e} + e \rightarrow W^-\rightarrow \bar{\nu_{\tau} + \tau}$ :    the distance will be ten times smaller and the  showering azimuth angular velocity will appear $ten$ times faster  than already fast EeV Tau Airshower. Their curvature in AMIGA may leave a clear imprint in dense array Cherenkov (and in mostly absent underground (3 m) muon one).

\section{Conclusions: UHE $\nu_{\tau}$ Astronomy spectroscopy}
The very exciting Neutrino Astronomy is beyond the corner. Either by EeV GZK \cite{Greisen:1966jv},\cite{za66} secondaries or by tens PeV signals
the detection via FD in AUGER is at the detection edge. The short and near air-shower from the telescope due to the lower energy makes the shower timing brief, sharp and at small zenith angle, and, upgoing.
On the contrary the unique inclined horizontal down-going hadron shower are far away, diluted in air density and in time scale as well as at possible bifurcate (by geomagnetic field) structure. No way to be confused.
Three or more times duration and morphology of \emph{up or down} signature will disentangle any rare neutrino lights arriving from their
unique but unusual sky: the Earth. Their rate, timing, energy and inclination may teach us on the real nature (nucleon or lighters nuclei) of
UHECR. In this hope we suggest (a) to implement the present AUGER array with additional array searching for shadows of inclined hadronic from Ande in AUGER; (b) To add Mini-Cherenkov telescope arrays facing Ande screen to reveal tau air-showers signals beyond the mountain chain also via direct Cherenkov flashes.(c) Locating a new telescope array few km distance to reduce the energy thresholds (d) Introducing novel trigger detection by fast horizontal track time signature.(e) Locating SD nearest to FD in order to enlarge their ability to discover the inclined events  (blazing Cherenkov flashes) both in FD and in Cherenkov lights as well as in muon tracks in nearby SD array elements. It must be remind that the AMIGA muon versus
SD Cherenkov signal may well signal the electromagnetic versus Muonic nature of horizontal air-shower, offering an additional tool to disentangle neutrino versus  hadronic air-showers. We foresee in very near years, possibly this and next one, the Tau Neutrino Astronomy birth, if accurate attention on trigger thresholds, and AMIGA-HEATS implementation will be concluded soon. \\



\end{document}